\documentclass[twocolumn]{aastex61}

\usepackage{url}
\usepackage{ulem}
\usepackage{CJKutf8}

\newcommand{\appropto}{\mathrel{\vcenter{\offinterlineskip\halign{\hfil$##$\cr\propto\cr\noalign{\kern2pt}\https://v2.overleaf.com/project/5b8956fe43cfc767092b2e0dsim\cr\noalign{\kern-2pt}}}}}
\newcommand{\bjdtdb}{\ensuremath{\rm {BJD_{TDB}}}}
\newcommand{\feh}{\ensuremath{\left[{\rm Fe}/{\rm H}\right]}}

\newcommand{\teff}{\ensuremath{T_{\rm eff}}}
\newcommand{\logg}{\ensuremath{\log{g_*}}}

\newcommand{\msun}{\ensuremath{\,M_\Sun}}
\newcommand{\rsun}{\ensuremath{\,R_\Sun}}
\newcommand{\lsun}{\ensuremath{\,L_\Sun}}
\newcommand{\mj}{\ensuremath{\,M_{\rm J}}}
\newcommand{\rj}{\ensuremath{\,R_{\rm J}}}
\newcommand{\fave}{\langle F \rangle}
\newcommand{\fluxcgs}{10$^9$ erg s$^{-1}$ cm$^{-2}$}

\newcommand{\figr}[1]{Fig.~\ref{fig:#1}}
\newcommand{\secr}[1]{Sec.~\ref{sec:#1}}

\newcommand{\tabr}[1]{\mbox{Table~\ref{tab:#1}}}

\newcommand{\targetsys}{HD\,202772}
\newcommand{\target}{\targetsys{}A}
\newcommand{\targetb}{\target{}\,b}
\newcommand{\companion}{\targetsys{}B}

\shorttitle{HD~202772A~\MakeLowercase{b}}
\shortauthors{Wang et al.}

\begin{document}

\title{HD~202772A~\MakeLowercase{b}: A Transiting Hot Jupiter Around a Bright, Mildly Evolved Star in a Visual Binary Discovered by \textit{TESS}}  




\author{Songhu Wang} 
\affiliation{Department of Astronomy, Yale University, New Haven, CT 06511, USA}
\affiliation{\textit{51 Pegasi b} Fellow}

\author{Matias Jones} 
\affiliation{European Southern Observatory, Casilla 19001, Santiago, Chile}

\author{Avi Shporer} 
\affiliation{Department of Physics and Kavli Institute for Astrophysics and Space Research, Massachusetts Institute of Technology, Cambridge, MA 02139, USA}

\author{Benjamin J.~Fulton} 
\affiliation{California Institute of Technology, Pasadena, CA 91125, USA}
\affiliation{IPAC-NASA Exoplanet Science Institute Pasadena, CA 91125, USA}

\author{Leonardo A. Paredes} 
\affiliation{Physics and Astronomy Department, Georgia State University, Atlanta, GA 30302, USA}

\author{Trifon Trifonov} 
\affiliation{Max-Planck-Institut f\"{u}r Astronomie, K\"{o}nigstuhl  17, 69117 Heidelberg, Germany}

\author{Diana Kossakowski} 
\affiliation{Max-Planck-Institut f\"{u}r Astronomie, K\"{o}nigstuhl  17, 69117 Heidelberg, Germany}

\author{Jason Eastman} 
\affiliation{Harvard-Smithsonian Center for Astrophysics, 60 Garden Street, Cambridge, MA 02138, USA}

\author{Maximilian N.\ G{\"u}nther} 
\affiliation{Department of Physics and Kavli Institute for Astrophysics and Space Research, Massachusetts Institute of Technology, Cambridge, MA 02139, USA}
\affiliation{Juan Carlos Torres Fellow}

\author{Chelsea X.~Huang} 
\affiliation{Department of Physics and Kavli Institute for Astrophysics and Space Research, Massachusetts Institute of Technology, Cambridge, MA 02139, USA}
\affiliation{Juan Carlos Torres Fellow}

\author{Sarah Millholland} 
\affiliation{Department of Astronomy, Yale University, New Haven, CT 06511, USA}
\affiliation{NSF Graduate Research Fellow}

\author{Darryl Seligman} 
\affiliation{Department of Astronomy, Yale University, New Haven, CT 06511, USA}

\author{Debra Fischer} 
\affiliation{Department of Astronomy, Yale University, New Haven, CT 06511, USA}

\author{Rafael Brahm} 
\affiliation{Center of Astro-Engineering UC, Pontificia Universidad Cat\'olica de Chile, Av. Vicu\~{n}a Mackenna 4860, 7820436 Macul, Santiago, Chile}
\affiliation{Instituto de Astrof\'isica, Pontificia Universidad Cat\'olica de Chile, Av.\ Vicu\~na Mackenna 4860, Macul, Santiago, Chile}
\affiliation{Millennium Institute for Astrophysics, Chile}

\author{Xian-Yu Wang} 
\affiliation{Key Laboratory of Optical Astronomy, National Astronomical Observatories, Chinese Academy of Sciences, Beijing 100012, China}
\affiliation{University of Chinese Academy of Sciences, Beijing, 100049, China}

\author{Bryndis Cruz} 
\affiliation{Department of Astronomy, Yale University, New Haven, CT 06511, USA}

\author{Hodari-Sadiki James} 
\affiliation{Physics and Astronomy Department, Georgia State University, Atlanta, GA 30302, USA}

\author{Brett Addison} 
\affiliation{University of Southern Queensland, Toowoomba, Qld 4350, Australia}

\author{Todd Henry} 
\affiliation{RECONS Institute, Chambersburg, PA, USA}

\author{En-Si Liang} 
\affiliation{School of Astronomy and Space Science \& Key Laboratory of Modern Astronomy and Astrophysics in Ministry of Education, Nanjing University, Nanjing 210023, China}

\author{Allen B. Davis} 
\affiliation{Department of Astronomy, Yale University, New Haven, CT 06511, USA}

\author{Ren\'{e} Tronsgaard} 
\affiliation{DTU Space, National Space Institute, Technical University of Denmark, Elektrovej 328, DK-2800 Kgs. Lyngby, Denmark}

\author{Keduse Worku} 
\affiliation{Department of Astronomy, Yale University, New Haven, CT 06511, USA}

\author{John Brewer} 
\affiliation{Department of Astronomy, Yale University, New Haven, CT 06511, USA}

\author{Martin K\"{u}rster} 
\affiliation{Max-Planck-Institut f\"{u}r Astronomie, K\"{o}nigstuhl  17, 69117 Heidelberg, Germany}


\author{Charles A.~Beichman} 
\affiliation{IPAC-NASA Exoplanet Science Institute Pasadena, CA 91125, USA}

\author{Allyson Bieryla}
\affiliation{Harvard-Smithsonian Center for Astrophysics, 60 Garden Street, Cambridge, MA 02138, USA}

\author{Timothy M.~Brown} 
\affiliation{Las Cumbres Observatory, 6740 Cortona Dr., Suite 102, Goleta, CA 93117, USA}
\affiliation{University of Colorado/CASA, Boulder, CO 80309, USA.}

\author{Jessie L.~Christiansen} 
\affiliation{IPAC-NASA Exoplanet Science Institute Pasadena, CA 91125, USA}

\author{David R.~Ciardi} 
\affiliation{IPAC-NASA Exoplanet Science Institute Pasadena, CA 91125, USA}

\author{Karen A.~Collins}
\affiliation{Harvard-Smithsonian Center for Astrophysics, 60 Garden Street, Cambridge, MA 02138, USA}

\author{Gilbert A.~Esquerdo}
\affiliation{Harvard-Smithsonian Center for Astrophysics, 60 Garden Street, Cambridge, MA 02138, USA}

\author{Andrew W. Howard} 
\affiliation{California Institute of Technology, Pasadena, CA 91125, USA}

\author{Howard Isaacson} 
\affiliation{Astronomy Department, University of California, Berkeley, CA 94720, USA}

\author{David W.~Latham}
\affiliation{Harvard-Smithsonian Center for Astrophysics, 60 Garden Street, Cambridge, MA 02138, USA}

\author{Tsevi Mazeh} 
\affiliation{School of Physics and Astronomy, Tel Aviv University, Tel Aviv 69978, Israel}

\author{Erik A. Petigura}
\affiliation{California Institute of Technology, Pasadena, CA 91125, USA}

\author{Samuel N.~Quinn} 
\affiliation{Harvard-Smithsonian Center for Astrophysics, 60 Garden Street, Cambridge, MA 02138, USA}

\author{Sahar Shahaf} 
\affiliation{School of Physics and Astronomy, Tel Aviv University, Tel Aviv 69978, Israel}

\author{Robert J.~Siverd} 
\affiliation{Department of Physics and Astronomy, Vanderbilt University, Nashville, TN 37235, USA}



\author{George R.~Ricker}	 
\affiliation{Department of Physics and Kavli Institute for Astrophysics and Space Research, Massachusetts Institute of Technology, Cambridge, MA 02139, USA}

\author{Roland Vanderspek} 
\affiliation{Department of Physics and Kavli Institute for Astrophysics and Space Research, Massachusetts Institute of Technology, Cambridge, MA 02139, USA}


\author{Sara Seager}
\affiliation{Department of Physics and Kavli Institute for Astrophysics and Space Research, Massachusetts Institute of Technology, Cambridge, MA 02139, USA}
\affiliation{Earth and Planetary Sciences, MIT, 77 Massachusetts Avenue, Cambridge, MA 02139, USA}

\author{Joshua N.~Winn}
\affiliation{Department of Astrophysical Sciences, Princeton University, 4 Ivy Lane, Princeton, NJ 08544, USA}

\author{Jon M.~Jenkins}
\affiliation{NASA Ames Research Center, Moffett Field, CA 94035, USA}

\author{Patricia T. Boyd}
\affiliation{NASA Goddard Space Flight Center, 8800 Greenbelt Road, Greenbelt, MD 20771, USA}

\author{G\'{a}bor F\H{u}r\'{e}sz}
\affiliation{Department of Physics and Kavli Institute for Astrophysics and Space Research, Massachusetts Institute of Technology, Cambridge, MA 02139, USA}

\author{Christopher Henze} 
\affiliation{NASA Ames Research Center, Moffett Field, CA 94035, USA}

\author{Alen M.~Levine}
\affiliation{Department of Physics and Kavli Institute for Astrophysics and Space Research, Massachusetts Institute of Technology, Cambridge, MA 02139, USA}

\author{Robert Morris} 
\affiliation{SETI Institute}  

\author{Martin Paegert}
\affiliation{Harvard-Smithsonian Center for Astrophysics, 60 Garden Street, Cambridge, MA 02138, USA}

\author{Keivan G. Stassun } 
\affiliation{Department of Physics and Astronomy, Vanderbilt University, Nashville, TN 37235, USA}
\affiliation{Department of Physics, Fisk University, Nashville, TN 37208, USA}

\author{Eric B.~Ting} 
\affiliation{NASA Ames Research Center, Moffett Field, CA 94035, USA}

\author{Michael Vezie}
\affiliation{Department of Physics and Kavli Institute for Astrophysics and Space Research, Massachusetts Institute of Technology, Cambridge, MA 02139, USA}


\author{Gregory Laughlin} 
\affiliation{Department of Astronomy, Yale University, New Haven, CT 06511, USA}

\correspondingauthor{Songhu Wang}
\email{song-hu.wang@yale.edu}

\begin{abstract} 
We report the first confirmation of a hot Jupiter discovered by the \textit{Transiting Exoplanet Survey Satellite} (\textit{TESS}) mission: \targetb.  The transit signal was detected in the data from \textit{TESS} Sector 1, and was confirmed to be of
planetary origin through radial velocity (RV) measurements.
\targetb\ is orbiting a mildly evolved star with a period of $3.3\,{\rm days}$. 
With an apparent magnitude of $V=8.3$, the star is among the brightest
known to host a hot Jupiter.
Based on the $27\,{\rm days}$ of \textit{TESS} photometry, and RV data from the CHIRON and HARPS spectrographs, the planet has a mass of 
 $1.008^{+0.074}_{-0.079}\,{\rm M_{J}}$ and radius of $1.562^{+0.053}_{-0.069}\,{\rm R_{J}}$, making it an inflated gas giant.
\targetb\ is a rare example of a transiting hot Jupiter around a
quickly evolving star.  It is also one of the most strongly irradiated hot Jupiters currently known.
\end{abstract} 


\section{Introduction}

Hot Jupiters, owing to their ease of detectability, are the best-studied population
of extrasolar planets.
However, we still do not understand how these behemoths came into existence. Did they form \textit{in situ} \citep{Bodenheimer2000, Batygin2016}, or did they arise in
wider orbits and migrate to their current locations \citep{Lin1996}?
If hot Jupiters did undergo migration, was this process violent \citep{Wu2007, Rasio1996, Wu2011, Petrovich2015} or quiescent \citep{Lin1996}?
Are the highly inclined and eccentric orbits of some hot Jupiters a consequence of high-eccentricity migration \citep{Winn2010, Bonomo2017}, or other mechanisms that are unrelated to planet migration \citep{Lai2016, Duffell2015}?
What is the occurrence rate of hot Jupiters as a function of stellar age \citep{Donati2016}?
What is the meaning of the high rate of distant companions \citep{knutson2014} and
the low rate of close-in companions \citep{Becker2015} to hot Jupiters?
What are the connections between hot Jupiters and warm Jupiters \citep{Huang2016}, hot
Neptunes \citep{Dong2018}, compact multiple-planet systems \citep{Lee2016}, and ultra-short-period planets \citep{Winn2018}?  Answers to these questions
may come more easily if we enlarge the sample of hot Jupiters around
very bright stars, subject to a wide range of irradiation levels.

The recently commissioned {\it Transiting Exoplanet Survey Satellite} (\textit{TESS}; \citealt{Ricker2015}) mission has the main goal of
discovering transiting exoplanets around bright and nearby stars, thereby facilitating
follow-up studies.
A few dozen hot Jupiters orbiting bright ($V\lesssim 10$ mag) stars are expected to emerge from the \textit{TESS} mission \citep{Sullivan2015, Barclay2018, Huang2018}.
They will always be among the most observationally favorable transiting planets,
and as such, they will observed
and re-observed in perpetuity as astronomical capabilities advance.

Here we report the first confirmation of a hot Jupiter discovered by the \textit{TESS} mission, \targetb{}. 
\secr{data} presents the data. \secr{stellar} describes
the derivation of the host star characteristics, and \secr{global} presents the system
parameters based on fitting the available photometry and RV data.
\secr{dis} summarizes the results and places this discovery into context.


\begin{deluxetable*}{l l l l}[bt]
\hspace{-1in}\tabletypesize{\scriptsize}
\tablecaption{\label{tab:target} \targetsys{}}
\tablewidth{0pt}
\tablehead{
\colhead{Parameter}  & \colhead{\target} & \colhead{\companion} & \colhead{Source}
}
\startdata
R.A. (hh:mm:ss)                &  21:18:47.901            &  21:18:47.813           & Gaia DR2 \\  
Dec. (dd:mm:ss)                & $-$26:36:58.95             &  $-$26:36:58.42           & Gaia DR2 \\   
$\mu_{\alpha}$ (mas~yr$^{-1}$) &  28.360 $\pm$ 0.269        &   23.236	$\pm$ 0.157	 & Gaia DR2 \\
$\mu_{\delta}$ (mas~yr$^{-1}$) &  $-$56.533 $\pm$ 0.418       & $-$57.557	$\pm$ 0.152    & Gaia DR2  \\
Parallax (mas)                 &  6.166 $\pm$ 0.092       & 6.686 $\pm$ 0.109         & Gaia DR2 \\
$B$ (mag)                      &  8.81 $\pm$ 0.02         &     10.65 $\pm$ 0.02    & Tycho\\
$V$ (mag)                      &  8.320 $\pm$ 0.05        &    10.15 $\pm$ 0.05     & Tycho\\
\textit{TESS} (mag)            &  7.92 $\pm$ 0.09           &     9.62 $\pm$ 0.09       & TIC V7$^1$\\
$J$ (mag)                     &  7.437 $\pm$ 0.027           & 9.142 $\pm$ 0.029      & NIRC2; this paper\\
$H$ (mag)                      & 7.266 $\pm$ 0.021           & 8.897 $\pm$ 0.022      & NIRC2; this paper\\
$Ks$ (mag)                     & 7.149 $\pm$ 0.017            & 8.858 $\pm$ 0.018     & NIRC2; this paper\\
\multicolumn{4}{l}{\hspace{1cm}\em Spectroscopic and Derived Properties} \\
\teff\ (K)              &  6330	$\pm$ 100           & 6156 $\pm$ 100            & Keck/HIRES; this paper\\
\logg\ (cgs)                &  4.03  $\pm$ 0.10          & 4.24 $\pm$ 0.10           & Keck/HIRES; this paper\\
\feh\ (dex)           &  0.29  $\pm$ 0.06          & 0.25 $\pm$ 0.06           & Keck/HIRES; this paper\\
$M_*$ ($\msun$)                  &  $1.69^{+0.05}_{-0.04}$   & 1.21 $\pm$ 0.04           & Keck/HIRES; this paper\\
$R_*$ ($\rsun$)                  &  $2.515^{+0.137}_{-0.127}$ & 1.16 $\pm$ 0.06           & Keck/HIRES; this paper\\
$Age$ (age)                    &  $1.52^{+0.19}_{-0.20}$    & $1.27^{+1.32}_{-0.80}$    & Keck/HIRES; this paper\\
\enddata
\tablenotetext{1}{\cite{Stassun2018}}
\end{deluxetable*}

\begin{deluxetable*}{l l l l l l}[bt]
\hspace{-1in}\tabletypesize{\scriptsize}
\tablecaption{\label{tab:stellarparams} \target}
\tablewidth{0pt}
\tablehead{
\colhead{Parameter}  & \colhead{SMARTS 1.5m/CHIRON} & \colhead{FLWO 1.5m/TRES} & \colhead{LCO/NRES} & \colhead{KECK/HIRES}& \colhead{EXOFASTv2 FIT}
}
\startdata
\teff\ [K]       &  6470 $\pm$ 100   &  6270 $\pm$ 50     &   6255	$\pm$ 100       	& 6330	$\pm$ 100            & $6230^{+110}_{-98}$     \\
\logg\ [cgs]         &  3.90  $\pm$ 0.15   &  3.91 $\pm$ 0.10   &   4.0	$\pm$ 0.1           & 4.03 $\pm$ 0.10            & $3.835 \pm 0.034$ \\
\feh\ [dex]    &  0.30  $\pm$ 0.10  &  0.16 $\pm$ 0.08   &   0.27 $\pm$ 0.06           & 0.29 $\pm$ 0.06            & $0.29^{+0.13}_{-0.24}$ \\
$M_*$ [$\msun$]         &  1.73 $\pm$ 0.05  &       ...          &   $1.78^{+0.02}_{-0.06}$    & $1.69^{+0.05}_{-0.04}$      & $1.703^{+0.075}_{-0.12}$ \\  
$R_*$ [$\rsun$]         &  2.65 $\pm$ 0.15  &       ...          &   $2.87^{+0.11}_{-0.10}$    & $2.515^{+0.137}_{-0.127}$   & $2.614^{+0.08}_{-0.11}$ \\
Age [Gyr]               &  1.6 $\pm$ 0.1    &       ...          &   $1.48^{+0.24}_{-0.16}$    & $1.52^{+0.19}_{-0.20}$      & $1.80^{+0.43}_{-0.30}$ \\
$V$~sin~$i$ [km\,s$^{-1}$]&   ...             &   8.1 $\pm$ 0.5    &       5.5 $\pm$ 1.4         &    7.0 $\pm$ 1.0            &        ...             \\ 
\enddata
\end{deluxetable*}

\section{Observation and Data Reduction}
\label{sec:data}

\subsection{\textit{TESS} Photometry
\label{ss:TESS Photometry}}

\begin{figure}
\vspace{0cm}\hspace{0cm}
\includegraphics[width=\columnwidth]{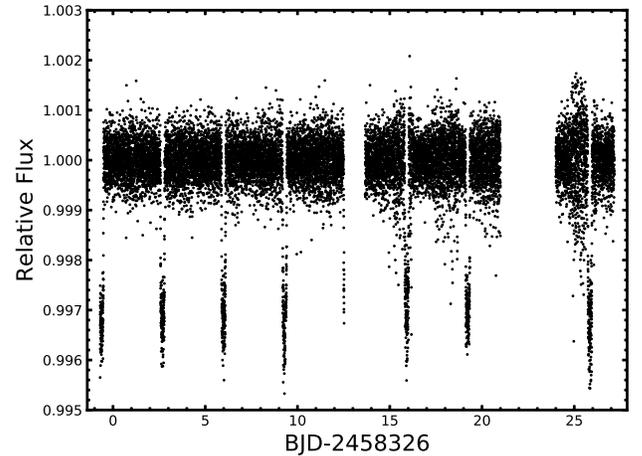}
\caption{
The \textit{TESS} Sector 1 light curve of \target, with two minute cadence.
Instrumental signals have already been removed from these data.
The periodic decreases in flux are the transits of \targetb.
The gap in the middle of the light curve is due to the data download,
which was performed at the end of the satellite's ninth orbit.
\label{fig:lc}}
\end{figure}

\targetsys\ (TIC~290131778, TOI~123) was observed by Camera 1 of
the \textit{TESS} spacecraft during the first sector of science operations,
between 2018 July 25 and 2018 August 22 (BJD 2458325 to 2458353).
The available data have two-minute time sampling (``short cadence''). Some basic parameters of the target are given in \tabr{target}.
Given its position in the sky, \targetsys\ will not be re-observed during the \textit{TESS} primary mission. The photometric data were analyzed by the Science Processing Operations Center (SPOC) pipeline, based on the NASA {\it Kepler} mission pipeline (Jenkins et al., \textit{in prep}). The light curve of \targetsys\ presented in \figr{lc} shows a clear transit signal. It was listed among the \textit{TESS} Alerts published online on 2018 September 5,
prompting us to download the photometric time series.\footnote{The \textit{TESS} Alerts are currently in a beta test phase. The full set of raw and calibrated data products from \textit{TESS} Sectors 1 through 4, including this source, will be available via NASA's Mikulski Archive for Space Telescopes (MAST) no later than January 2019.}

We detrended the raw light curves in the following way \citep[see e.g.][]{Guenther2017b, Guenther2018}.
After masking out all of the data obtained during transits,
we fitted a Gaussian Process (GP) model to the data, using a Matern 3/2 kernel and a white noise kernel.
For this task we employed the {\scshape celerite} package, which uses a Taylor-series expansion of these kernel functions. Once the parameters of the GP were constrained based on the out-of-transit data,
we used it to detrend the entire light curve.

The SPOC pipeline produces flags for poor-quality exposures. These include exposures taken during the 30-minute long momentum dumps that occurred every 2.5 days (10 times in total).
All flagged exposures were omitted from our analysis. The resulting light curve is plotted in \figr{lc}.

\begin{figure}
\vspace{0cm}\hspace{0cm}
\includegraphics[scale=0.35, angle=90]{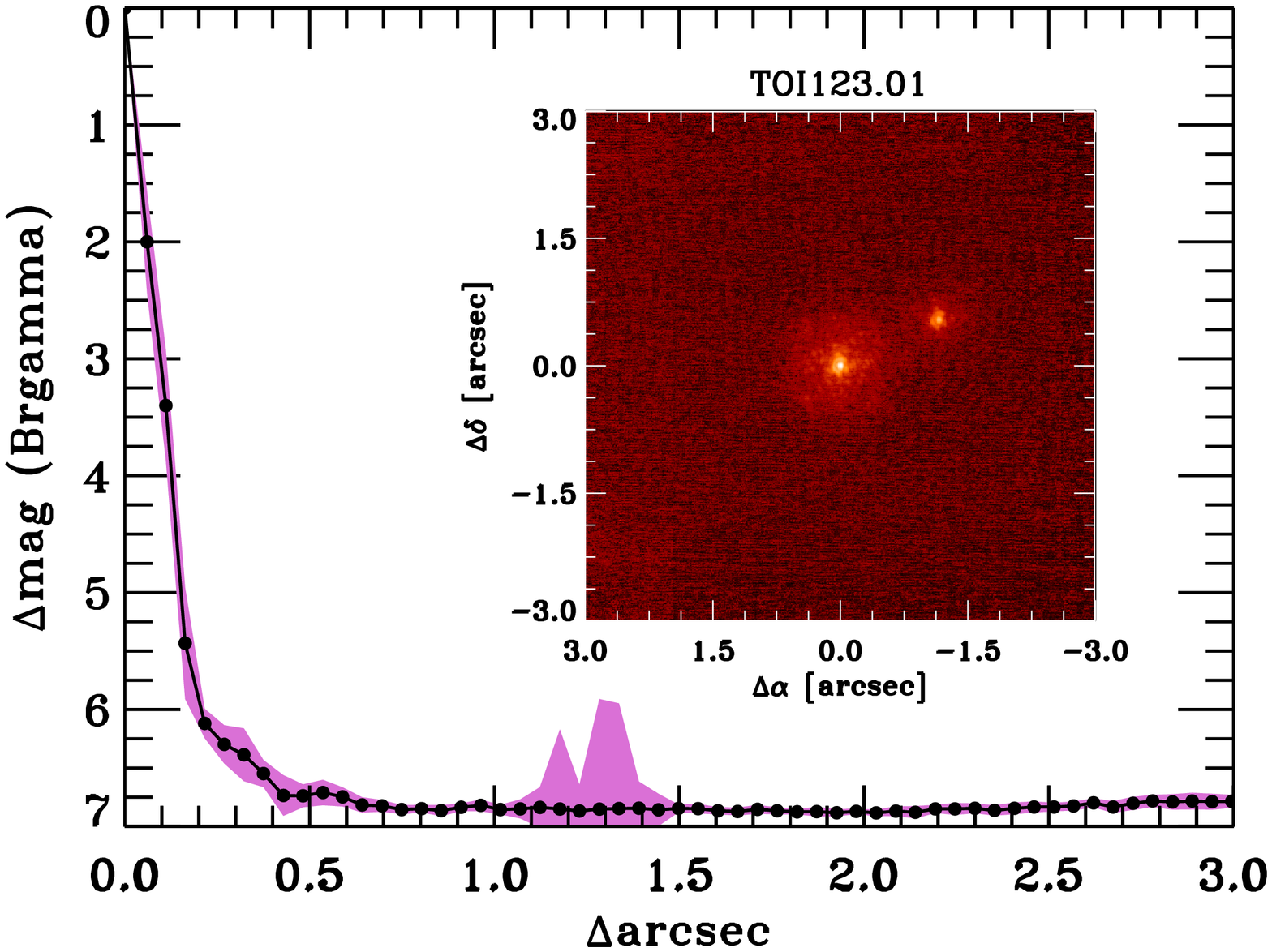}
\caption{AO image (inset) and $K_s$-band contrast curves for \target, obtained with Keck/NIRC2. A companion is visible 1.3\arcsec\ northwest of the primary. The black line is the $5\sigma$ sensitivity, with a 1$\sigma$ scatter marked in purple. See text for further details. 
\label{fig:ao_contrast}}
\end{figure}

\subsection{Keck/NIRC2 Adaptive Optics Imaging
\label{sec:AO}}

\targetsys\ was reported to be a pair of stars in several wide-field surveys (e.g.\ Tycho-2, \citealt{Hog2000}; PPMXL, \citealt{Roeser2010}; Gaia DR2, \citealt{gaia2018}; see also \citealt{Holden1978},
\citealt{Horch2001}) reported two bright stars separated by $\approx$1.5\arcsec, although the {\it Gaia} DR 2 catalog flags the brighter star as a ``duplicate'' entry.
To check on these earlier findings, we performed high-resolution
adaptive optics (AO) imaging at Keck Observatory.

The Keck observations were made with the NIRC2 instrument on Keck-II behind the natural guide star AO system. The observations were made on 2018 September 18 on a night with partial cirrus conditions.
We used the standard 3-point dither pattern that avoids
the left lower quadrant of the detector (which is typically noisier than the other three quadrants).
The dither pattern step size was 3\arcsec, and it was repeated twice, with the second dither offset from the first dither by 0.5\arcsec.  
Observations were made with three different filters: narrow-band Br$_{\gamma}$ $(\lambda_o = 2.1686; \Delta\lambda = 0.0326\,\mu$m), H-continuum $(\lambda_o = 1.5804; \Delta\lambda = 0.0232\,\mu$m), and J-continuum $(\lambda_o = 1.2132; \Delta\lambda = 0.0198\,\mu$m), using integration times of 1.45, 5.0, and 1.5 seconds, respectively.
The camera was in the narrow-angle mode with a full field of view of 10\arcsec\ and a pixel scale of approximately 0.01\arcsec\ per pixel.

Two stars were clearly detected, with a separation of 1.3\arcsec{}\ (Figure \ref{fig:ao_contrast}). The resolution of the 2\,\micron\ image is approximately 0.05\arcsec\ FWHM.
The sensitivity of the final combined AO image was determined by injecting simulated sources azimuthally around the primary target every $45^\circ$ at separations of integer multiples of the FWHM of the central source \citep{furlan2017}. The brightness of each injected source was scaled until standard aperture photometry detected it with $5\sigma$ significance. The resulting brightness of the injected sources relative to our target was taken
to be the contrast limit for the injected location. The final $5\sigma$ limit at each separation was determined from the average of all of the determined limits at that separation, with an uncertainty given by the RMS dispersion of the results for different azimuthal slices.
Figure \ref{fig:ao_contrast} shows the 2\,\micron\ sensitivity curve in black, with the $1\sigma$ (RMS) dispersion marked in purple. The
inset image shows the primary target in the center and the second source located 1.3\arcsec\ to the northwest.

The two stars were detected in all three filters. The presence of the blended companion must be taken into account to obtain the correct transit depth and planetary radius \citep{ciardi2015}. The stars have blended 2MASS magnitudes of $J = 7.232 \pm 0.026$ mag, $H=7.048 \pm 0.021$ mag, and $K_s = 6.945 \pm 0.026$ mag. The stars have measured magnitude differences of $\Delta J = 1.705 \pm 0.0015$ mag, $\Delta H = 1.631 \pm 0.008$ mag, and $\Delta K_s = 1.709 \pm 0.011$ mag.  The primary star has deblended apparent magnitudes of $J_1 = 7.437 \pm 0.027$ mag, $H_1 = 7.266 \pm 0.021$ mag, and $Ks_1 = 7.149 \pm 0.017$ mag, corresponding to $(J-H)_1 = 0.171 \pm 0.034$ mag and $(H-K_s)_1 = 0.117 \pm 0.027$ mag.
The secondary star has deblended apparent magnitudes of $J_2 = 9.142 \pm 0.029$ mag, $H_2 = 8.897 \pm 0.022$ mag, and $Ks_2 = 8.858 \pm 0.018$ mag, corresponding to $(J-H)_2 = 0.245 \pm 0.036$ mag and $(H-K_s)_2 = 0.039 \pm 0.029$ mag.
The infrared colors of the primary star are consistent with an early-G or late-F main sequence star, in agreement with the derived stellar parameters.  The companion star has infrared colors that are consistent with a later G-type main sequence star. 

Based on the \textit{TESS} magnitude and 2MASS color relationships established for the \textit{TESS} Input Catalog \citep{Stassun2018}, we estimate that the deblended \textit{TESS} magnitudes for the two components to be $T_{\rm 1} = 7.92 \pm 0.09$ mag and $T_{\rm 2} = 9.62 \pm 0.09$ mag for a \textit{TESS} magnitude difference of $\Delta_T = 1.7 \pm 0.1$ mag and a \textit{TESS} flux ratio of $F_{2}/F_{1}=0.21 \pm 0.02$.  We used this value of the flux ratio
to correct the apparent transit depth in the \textit{TESS} light curve, and derive the unblended transit depth.

We will refer to the brighter target (hosting the planet) as \target{}, and the fainter companion as \companion{}. 
Given the similarity between the two stars, and a projected separation of only $\sim 200$\,AU,
it seems very likely that the two stars are gravitationally bound.
The chance alignment probability is negligible, as estimated from the Besançon Galactic Model \citep{Robin2003}. 
We have also checked all 131 stars in the Gaia DR2 catalog within 300 \arcsec\ of \target\ with measured parallax and proper motion,
and the only nearby source with a similar projected velocity is \companion{}.

\subsection{LCO/NRES Optical Spectroscopy
\label{sec:Lco_spec}}

To obtain independent estimates of the stellar parameters, we performed high-resolution
optical spectroscopy with the Las Cumbres Observatory (LCO) robotic network of telescopes \citep{Brown2013}.
We obtained three 20-minute exposures with a total signal-to-noise ratio (SNR)~$\approx$100 with the Network of Echelle Spectrographs (NRES; \citealt{Siverd2016, Siverd2018}) mounted on a 1.0m telescope at the South African Astronomical Observatory (SAAO). 

Since the NRES fiber diameter corresponds to 2.8\arcsec, it captured the light from both stars in the visual binary system.
Using TODCOR \citep{Zucker1994}, we identified two RV components separated by $5 \pm 1$~km~s$^{-1}$.
This RV difference is compatible with the order of magnitude of the RV variation one would expect from
the orbital motion of the stars, given their masses and sky-projected separation.

\begin{figure}
\vspace{0cm}\hspace{0cm}
\includegraphics[width=\columnwidth]{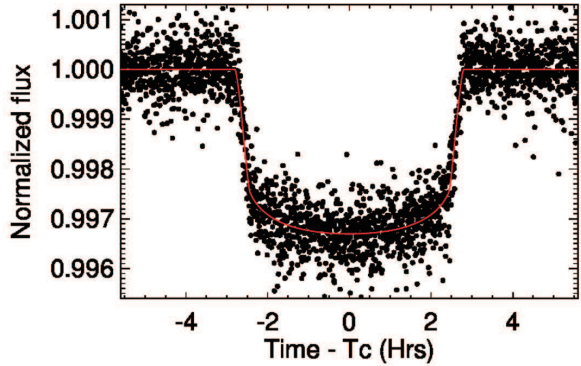}
\caption{Phased light curves of \targetb{}. 
The red solid line represents the best-fitting model.
\label{fig:phaselc}}
\end{figure}

\subsection{Keck/HIRES Optical Spectroscopy
\label{sec:Keck_spec}}


In order to obtain a spectrum of each of the two stars with minimal contamination from the other star, we observed both stars with Keck/HIRES \citep{Vogt1994} on 2018 September 23. We obtained one spectrum of each star, with the HIRES slit oriented perpendicular to the separation between the two stars.
Given the angular separation between the two stars, the slit width (0.86\arcsec) and the astronomical seeing at Keck at the night of the observation ($\approx$0.7\arcsec), the level of cross-contamination is expected to be less than 10\%.
Both spectra were obtained without the iodine (I$_2$) cell, at a spectroscopic resolution of R$\approx$65,000, and at a signal-to-noise ratio per pixel of 150 at 5500 \AA.
A similar technique was successfully applied by \cite{Shporer2014} to a visual binary system in which
both members have similar brightness and a smaller angular separation.

\begin{figure}
\vspace{0cm}\hspace{0cm}
\includegraphics[width=\columnwidth]{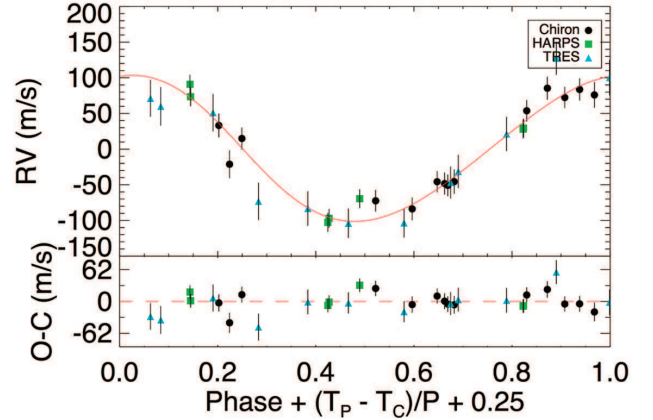}
\caption{RV measurements
from CHIRON (black circles), HARPS (green squares), and TRES (blue triangles) as a function of orbital phase. The error bars include the fitted jitter term. The units of the horizontal axis are chosen so the time of transit is at 0.25.
The solid red line is the best-fitting model, based on the transit photometry as well as the RV data.
The fitted value of the systemic velocity
has been subtracted from both the RVs and the model.
The bottom panel presents the residuals between the data and the best fit model. 
\label{fig:rvs}}
\end{figure}

\subsection{Doppler Velocimetry with CHIRON 
\label{sec:RV}}

We obtained a total of 14 spectra of \target\ using CHIRON \citep{Tokovinin2013}, a fiber-fed high-resolution optical spectrograph mounted on the SMARTS 1.5m telescope at Cerro Tololo in Chile. We collected the spectra using the image slicer, which delivers a resolution of $\sim$\,80,000 and a higher throughput than the standard slit mode. Our 15-minute exposures yielded a SNR per pixel of $\sim$\,60-80 at 5500\,\AA.

Although CHIRON is equipped with an iodine cell to obtain a precise wavelength solution that permits long-term RV precision better than $\sim$ 2-3 m\,s$^{-1}$ (e.g. \citealt{Jones2017}), we did not use the iodine
cell for these observations. This is because the cell absorbs $\sim$ 25 \% of the light at 5500\AA, significantly decreasing the signal-to-noise ratio.
Moreover, using the I$_2$ cell requires a time-consuming
acquisition of a high-SNR template spectrum of the target star.

Instead, we derived the RVs using the Cross-Correlation-Function (CCF) method, in a manner
similar to \citet{Jones2017}. CHIRON is not equipped with a simultaneous calibration fiber. Instead,
we acquired a Th-Ar lamp exposure before and after each target exposure. 
We computed the wavelength solution for the target spectra by interpolating line positions for the lamps to match the temporal midpoint of each observation. We thereby achieved a RV stability of $\sim$ 5-6 m$\,$s$^{-1}$, which was verified with two RV standard stars observed nightly. The resulting RVs of \target\ are listed in \tabr{rvs}.
RVs collected by CHIRON show a $\sim 95\,{\rm m\,s^{-1}}$ sinusoidal variation in phase with the transit ephemeris. 

Finally, from the CCF, we measured the bisector velocity span (BVS) and FWHM variations, to check
on the possibility that the observed RV variation
results from stellar activity or a background eclipsing binary system (see, e.g., \citealt{santerne2015}).
\figr{bvs} shows the BVS and FWHM as a function of the measured radial velocities.
There is no significant correlation between these quantities and the radial velocities. 

\begin{figure}
\vspace{0cm}\hspace{0cm}
\includegraphics[width=\columnwidth]{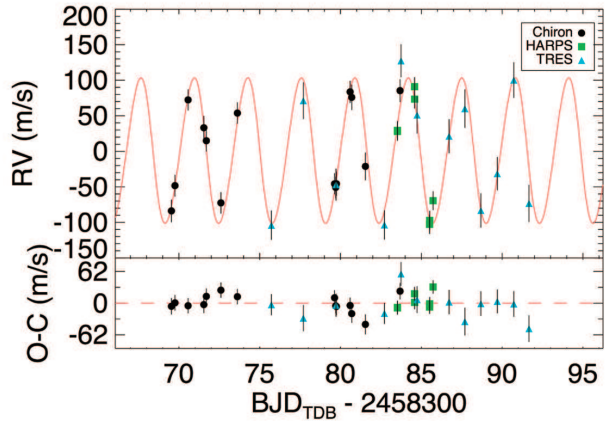}
\caption{Same as Figure \ref{fig:rvs}, but as a function
of time instead of orbital phase. 
\label{fig:rvunphased}}
\end{figure}

The CHIRON fiber has a 2.7\arcsec\ diameter on the sky, but \target\ and B are separated by only 1.3\arcsec{}, which means that we must expect some of the light from the binary companion to be present in the spectra.
Given that the two stars have a similar radial velocity, there is a risk that the stationary CCF of the binary companion causes the apparent amplitude of the RV variation to be lower than the true RV variation of the planet host.
Such a ``peak pulling'' effect was observed in a study of the Kepler-14 system \citep{Buchhave2011}.
We note, however, that \companion{} is fainter than \target\ and only emits about 20\% of the total light from the binary system. As described in Section~\ref{sec:harps_data}, we did not find any evidence that the RVs from CHIRON
were significantly affected by light contamination from \companion.

\subsection{Doppler Velocimetry with FLWO 1.5m/TRES
\label{sec:TRES_RV}}

We obtained 12 spectra of \target\ with the Tillinghast Reflector Echelle Spectrograph \citep[TRES;][]{furesz:2008} on the 1.5m Tillinghast Reflector at Fred L. Whipple Observatory (FLWO) on Mt. Hopkins, AZ between UT 2018 September 14 and UT 2018 September 30. TRES is a fiber-fed, cross-dispersed echelle spectrograph with a resolving power of $R\sim44,000$\ and an instrumental precision of $\sim10$--$15$\ m s$^{-1}$. The typical exposure time was $\sim4$\ minutes, resulting in SNR per resolution element of $\sim75$ at $5200$\,\AA. The spectra are calibrated using a ThAr lamp, exposed through the science fiber before and after each set of science exposures. We note that the TRES fiber is 2.3\arcsec, and the exposures therefore include light from \companion{}.

We reduced and analyzed the spectra according to the procedures outlined in \citet{buchhave2010}. Namely, the spectra were optimally extracted and then cross-correlated, order by order, against the strongest spectrum of \target. We exclude spectral orders far to the blue where the SNR is low, in the red where telluric lines contaminate the spectrum, and a few orders in between with little information content or affected by broad feature (e.g., Balmer lines) toward the edge of the order that affect continuum fitting. RVs were ultimately derived from a region spanning $4130$--$6280$\,\AA. The peak of the summed CCF across all orders is fit to derive the final RV, and the scatter between orders within a each spectrum is taken to be the internal error estimate. These relative RVs and their uncertainties are reported in Table~\ref{tab:rvs}. 

\begin{figure}
\vspace{0cm}\hspace{0cm}
\includegraphics[scale=0.5]{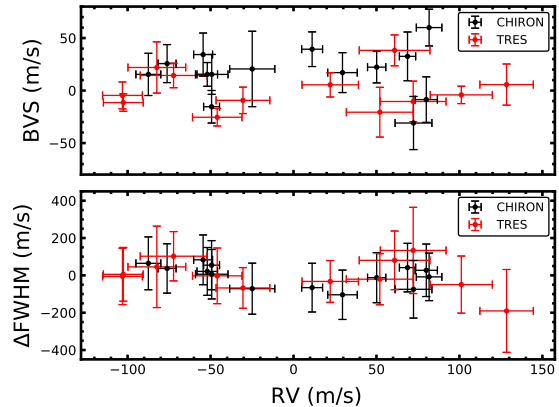}
\caption{Bisector velocity span (BVS; top panel) and CCF FWHM (bottom panel) as functions of the radial velocities collected by SMARTS 1.5m/CHIRON and FLWO 1.5m/TRES.
\label{fig:bvs}}
\end{figure}

We also derive the BVS and FWHM from the cross-correlation function of each spectrum against a non-rotating synthetic spectrum with appropriate \teff, \logg, and \feh. These values are also reported in Table~\ref{tab:rvs} and shown in Figure~\ref{fig:bvs}, and show no correlation with the RVs.

\subsection{Doppler Velocimetry with HARPS
\label{sec:harps_data}}

To provide further confirmation of the planetary origin of the transit signal,
we obtained 9 spectra using the High Accuracy Radial velocity Planet Searcher (HARPS; \citealt{harps}).
These data were obtained during three consecutive nights in good seeing conditions
($\lesssim$ 1.0\arcsec).
Exposure time was 200-300 seconds, leading to a SNR of $\sim$ 70-80 at 6,000 \AA. We carefully centered the brighter star within the 1\arcsec\ aperture fiber, to ensure that no light contamination from the companion was reaching the detector. We also carefully adjusted the size of the guiding box, to avoid guiding problems due to the secondary star, which was clearly visible in the acquisition camera. 
During the HARPS observations, the Moon was between 13$^\circ$ and 36$^\circ$ from our target, with an illuminated
fraction between 92\% and 99\%. This led to some lunar contamination in the spectra. Moreover, the RV of the Moon was very close to the RV of the target star, which severely affected the shape of the CCF.
For this reason, we discarded the two RV data points that were most affected and had very deviant values.
Also, due to the contamination, the derived BVS and FWHM of the CCF are not reliable, and are not listed in Table \ref{tab:rvs}.
We processed the HARPS data using the CERES code \citep{ceres}. The resulting RVs are listed in \tabr{rvs}, and also shown in Figure~\ref{fig:rvs}. As can be seen, the HARPS data agree with the CHIRON data, although the scatter around the fit is larger than expected. This is most likely caused by the lunar light contamination.

\section{Stellar Parameters}
\label{sec:stellar}

\subsection{Results from CHIRON}
\label{sec:chiron}

To derive the stellar atmospheric parameters of \target\ we measured the equivalent widths (EWs) of about 150 relatively weak Fe\,{\sc i} and Fe\,{\sc ii} absorption
lines (EW $\lesssim$ 120\,m\AA\,). The EWs were measured in the
high-SNR template obtained by stacking the individual spectra (see \secr{RV}), using the ARES\,v2 automatic tool \citep{Sousa2015}.

We then used the MOOG code \citep{Sneden1973} along with the \citet{Kurucz1993} stellar atmosphere models to solve the radiative transfer equations under the assumptions of local excitation and ionization equilibrium via the Saha and Boltzmann equations. For each iron line, MOOG computes the corresponding iron abundance by matching the measured EW in the curve of growth computed from the input stellar model. This procedure is performed iteratively for models with different effective temperatures (\teff), iron abundances (\feh), and micro-turbulent velocities ($V_{\rm micro}$) until there is no correlation between the line excitation potential and wavelength with the model abundance. Finally, we obtained the surface gravity (\logg) using the constraint that the iron abundances derived from both the Fe {\sc i} and Fe {\sc ii} lines should
be the same (For a more thorough description of the procedure, see \citealt{Jones2011}). Table~\ref{tab:stellarparams} gives the resulting stellar parameters.

We computed the luminosity of \target\ based on the
Gaia DR2 parallax ($\pi$ = 6.166 $\pm$ 0.092, \citealt{gaia2018}), the apparent $V$ magnitude (after correcting for interstellar absorption by A$_V$ = 0.10 mag),
and the bolometric correction of \citep{alonso1999}.
Using this information and the stellar atmospheric parameters (\teff\ and \feh), we derived the stellar physical parameters using the PARSEC stellar-evolutionary models \citep{bressan2012}. The results are also listed in Table~\ref{tab:stellarparams}.

\subsection{Results from FLWO 1.5m/TRES}
\label{sec:tres}

We used the Spectral Parameter Classification (SPC) tool \citep{buchhave2012} to derive stellar parameters from the TRES spectra. We allowed \teff, \logg, \feh, and $V\sin{i}$ to be free parameters. SPC works by cross correlating an observed spectrum against a grid of synthetic spectra based on Kurucz atmospheric models \citep{Kurucz1993}. The weighted average results are listed in Table~\ref{tab:stellarparams}.

\subsection{Results from LCO/NRES}
\label{sec:nres}

\begin{deluxetable}{cccccccc}
\tablewidth{0pt}
\tablecaption{\label{tab:rvs} Relative radial velocities for \target{}}
\tablehead{
\colhead{${\rm BJD}$} & \colhead{RV}                 & \colhead{$\sigma_{\rm RV}$}  &
\colhead{BVS$^1$}         & \colhead{$\sigma_{\rm BVS}$}   & \colhead{FWHM$^1$}             & \colhead{$\sigma_{\rm FWHM}$}  & \colhead{Instrument}\\
\colhead{-2458300}  & \colhead{m s$^{-1}$} & \colhead{m s$^{-1}$}             & 
\colhead{m s$^{-1}$}  & \colhead{m s$^{-1}$} & \colhead{m s$^{-1}$}             & \colhead{m s$^{-1}$}  
}
\startdata
69.5363 &  -87.5  &   7.5 &  15.4  &   20.1 &   16302.0 &    141.5 & CHIRON \\
69.7550 &  -52.0  &   6.2 &  15.4  &   11.3 &   16258.7 &    128.1 & CHIRON \\
70.5656 &   68.7  &   4.9 &  32.6  &   23.2 &   16279.1 &    130.0 & CHIRON \\
71.5411 &   29.5  &   8.9 &  17.1  &   19.0 &   16133.5 &    132.2 & CHIRON \\
71.6964 &   11.2  &   6.3 &  39.4  &   16.5 &   16171.8 &    130.5 & CHIRON \\
72.5988 &  -76.2  &   5.4 &  25.7  &   18.1 &   16274.6 &    132.2 & CHIRON \\
73.6179 &   50.1  &   5.5 &  22.3  &   15.0 &   16225.0 &    133.9 & CHIRON \\
79.6314 &  -49.4  &   4.8 & -15.4  &   15.5 &   16292.1 &    131.5 & CHIRON \\
79.7038 &  -54.5  &   5.7 &  34.3  &   20.5 &   16320.8 &    133.3 & CHIRON \\
79.7466 &  -49.3  &   9.8 &  15.4  &   19.0 &   16243.6 &    132.4 & CHIRON \\
80.5925 &   80.0  &   6.7 &  -8.6  &   21.7 &   16264.7 &    140.5 & CHIRON \\
80.6924 &   72.3  &  11.1 & -30.9  &   25.3 &   16162.8 &    154.1 & CHIRON \\
81.5400 &  -24.9  &  13.6 &  20.6  &   35.9 &   16166.2 &    136.7 & CHIRON \\
83.6838 &   81.7  &   7.7 &  60.0  &   17.5 &   16229.4 &    127.2 & CHIRON \\
83.5215  &  34.8  &   2.0 &  ...   &   ...  &   ...     &    ...   & HARPS  \\
83.5252  &  36.3  &   2.0 &  ...   &   ...  &   ...     &    ...   & HARPS  \\ 
84.5863  &  80.3  &   2.0 &  ...   &   ...  &   ...     &    ...   & HARPS  \\
84.5837  &  97.8  &   2.4 &  ...   &   ...  &   ...     &    ...   & HARPS  \\
85.5194  & -90.3  &   2.8 &  ...   &   ...  &   ...     &    ...   & HARPS  \\
85.5128  & -96.0  &   2.8 &  ...   &   ...  &   ...     &    ...   & HARPS  \\
85.7274  & -62.7  &   2.0 &  ...   &   ...  &   ...     &    ...   & HARPS  \\
75.7239  &  -0.5  &  12.0 &  -4.6  &   12.8 &   18858   &    151   & TRES  \\
77.6977  & 174.7  &  19.8 & -10.4  &   19.4 &   18997   &    231   & TRES  \\
79.7226  &  56.6  &  14.9 & -25.4  &    8.3 &   18861   &    149   & TRES  \\
82.7170  &   0.0  &  12.0 & -11.4  &    8.3 &   18869   &    144   & TRES  \\
83.7441  & 231.0  &  16.1 &   5.7  &   19.6 &   18673   &    222   & TRES  \\
84.7374  & 154.6  &  20.2 & -20.6  &   23.7 &   18843   &    137   & TRES  \\
86.7178  & 124.7  &  17.0 &   5.5  &   11.6 &   18831   &    112   & TRES  \\
87.6938  & 163.5  &  21.3 &  38.3  &   14.7 &   18944   &    157   & TRES  \\
88.6862  &  20.2  &  17.4 &  22.0  &   24.4 &   18909   &    218   & TRES  \\
89.7004  &  72.1  &  16.2 &  -9.3  &   12.6 &   18796   &    109   & TRES  \\
90.7214  & 203.7  &  18.7 &  -4.1  &    8.3 &   18814   &    153   & TRES  \\
91.6617  &  30.3  &  20.1 &  14.4  &   11.7 &   18966   &    132   & TRES  \\
\enddata
\tablenotetext{1}{The HARPS BVS and FWHM are not listed because those measurements
were corrupted by moonlight (Section \ref{sec:harps_data}).}
\end{deluxetable}

\begin{figure*}
\vspace{0cm}\hspace{0cm}
\includegraphics[width=\columnwidth]{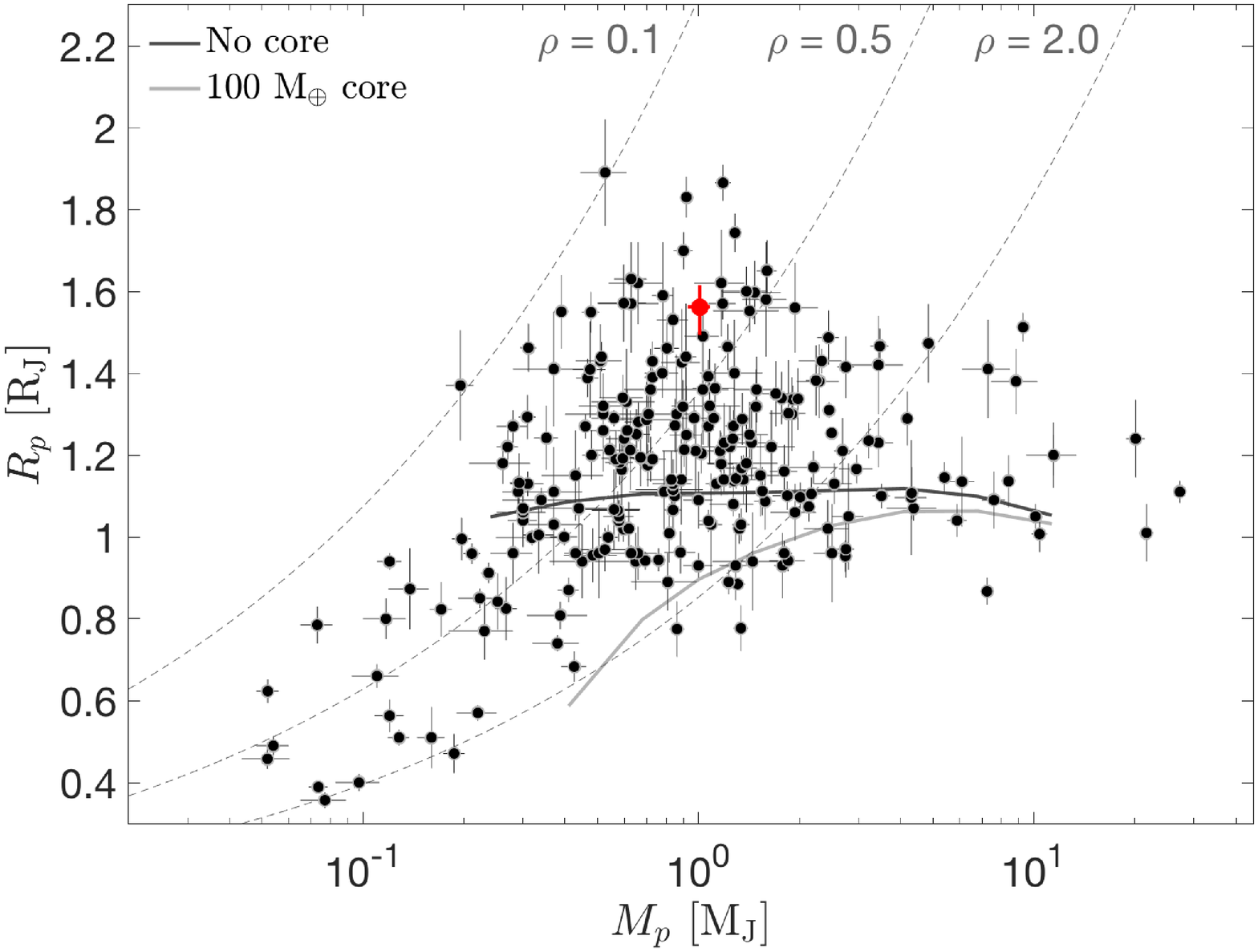}
\includegraphics[width=\columnwidth]{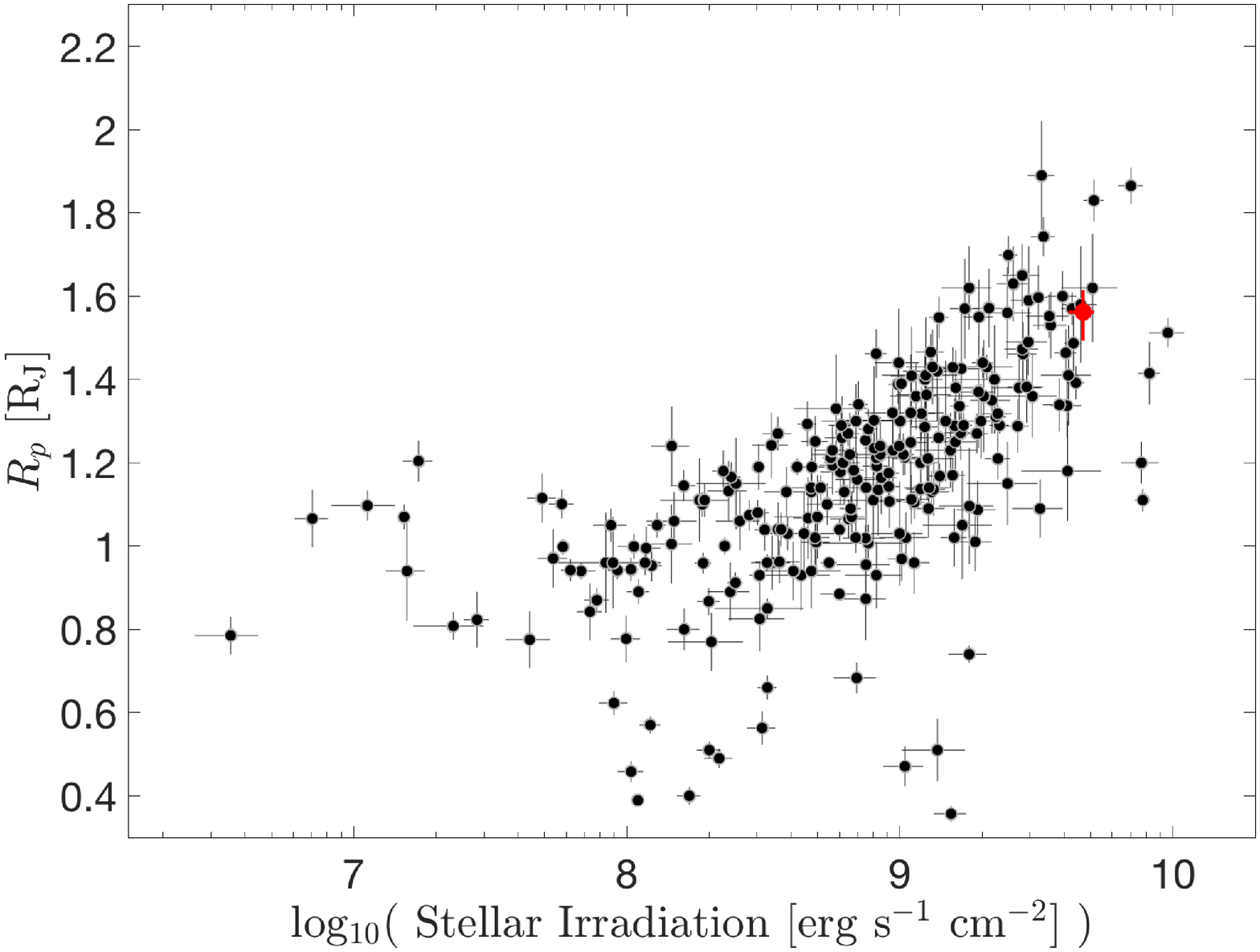}
\caption{The position of \targetb\ (red) in the space of
mass, radius, and irradiation, compared to the population of known transiting gas giant planets (black). In the left panel, the solid lines mark theoretical models taken from \cite{Baraffe2014} for no core (black)
and a 100~M$_{\rm Earth}$ core (gray).
The dashed lines are isodensity contours.
Data were obtained from the NASA Exoplanet Archive \citep{Akeson2013} on 2018 September 15.
\label{fig:radmassflux}}
\end{figure*}

We analyzed the LCO/NRES spectrum using the methodology of \citet{Fulton2018}. We measured \teff, \logg, \feh, and $V\sin{i}$ using \texttt{SpecMatch} \citep{Petigura2015}\footnote{https://github.com/petigura/specmatch-syn}, which compares the observed spectrum with a grid of model spectra \citep{Coelho2005}.
The resulting parameters are listed in Table~\ref{tab:stellarparams}.


To calculate the star's physical parameters,
we used \texttt{isoclassify} \citep{Huber2017}, which takes as input the
effective temperature, metallicity, parallax,
and apparent $K_{s}$ magnitude.  Using the \texttt{isoclassify} ``direct''
mode, we calculated the posterior probability distributions for $R_{\star}$ and $L_{\star}$ by applying the Stefan–Boltzmann law.  Using the \texttt{isoclassify} ``grid'' mode, we calculated the range of MIST isochrone models \citep{Dotter2016, Choi2016} that are consistent with the spectroscopic parameters to estimate the stellar mass and age. The results of the \texttt{SpecMatch}+\texttt{isoclassify} analysis are listed in Table~\ref{tab:stellarparams}.


\subsection{Results from Keck/HIRES}

The Keck/HIRES spectrum of each of the two stars was also
analyzed using \texttt{SpecMatch}.
The resulting atmospheric and physical stellar parameters for both \target\ and B are listed in Table~\ref{tab:target} and Table~\ref{tab:stellarparams}.

The stellar parameters derived from resolved HIRES spectra show good agreement with those from the CHIRON, TRES, and NRES spectra, except that TRES finds modestly lower \feh. 
Evidently, the contaminating light from the secondary star in the CHIRON, TRES, and NRES spectra did not strongly affect the determination of the basic stellar parameters.

\section{Planetary System Parameters from Global Analysis}
\label{sec:global}

We performed a joint analysis of the \textit{TESS} data, the RV data, and
the stellar spectral energy distribution 
using EXOFASTv2\footnote{https://github.com/jdeast/EXOFASTv2} \citep{Eastman2013, Eastman2017}.
The stellar limb darkening function was assumed to be quadratic, with the
coefficients fit with a prior from \citet{Claret2018} for the \textit{TESS} band based on the \logg, \teff, and \feh \ at each step. We imposed Gaussian priors on the Gaia DR2 parallax of $6.77 \pm 0.11$ mas (after adjusting by 82 $\mu as$ as
advocated by \citealt{Stassun2018}) and the \textit{TESS}-band dilution
from the neighboring star of $0.21 \pm 0.02$ found from the AO imaging. We imposed an upper limit on the $V$-band extinction of 0.17236 from \citet{Schlafly2011}.
The priors for all the remaining parameters were uniform and unbounded.

To constrain the spectral energy distribution, we use the broadband photometry from Tycho (which resolved the companion), and the 2MASS $JHK$ photometry after
deblending based on the Keck AO images.
We opted not to impose any informative priors on the spectroscopic parameters
\teff\ or \feh. Instead, we relied on the observed
spectral energy distribution and the MIST stellar-evolutionary models
to constrain the stellar parameters.
The resulting stellar parameters are listed in Table~\ref{tab:stellarparams}, and
show good agreement with the results of the spectroscopic analysis.
Table~\ref{tab:params} gives the results of the EXOFAST fit to all of the data.
The best-fitting model is also
plotted in Figures~\ref{fig:phaselc}, \ref{fig:rvs}, and \ref{fig:rvunphased}.

\section{Discussion}
\label{sec:dis}

\targetb\ is an inflated Jupiter-mass planet orbiting a metal-rich star with an orbital period of $3.3\,{\rm days}$. 
The red dots in \figr{radmassflux} show the location
of this newly discovered planet in the spaces of planetary
mass, radius, and incident flux, compared
with the current sample of transiting giant planets.
\targetb\ is one of the largest known planets,
with a relatively low mean density of 0.33 g cm$^{-3}$.
It is also one of the most strongly irradiated planets,
thereby obeying the well known correlation between
planetary radius and degree of irradiation
\citep[see, e.g.,][]{Laughlin2011, Lopez2016}.
Based on the irradiation of 4.7$\times$10$^{9}$ erg s$^{-1}$ cm$^{-2}$,
the estimated equilibrium temperature is about 2{,}100 K
(see \tabr{params}).

\begin{figure}
\vspace{0cm}\hspace{0cm}
\includegraphics[width=\columnwidth]{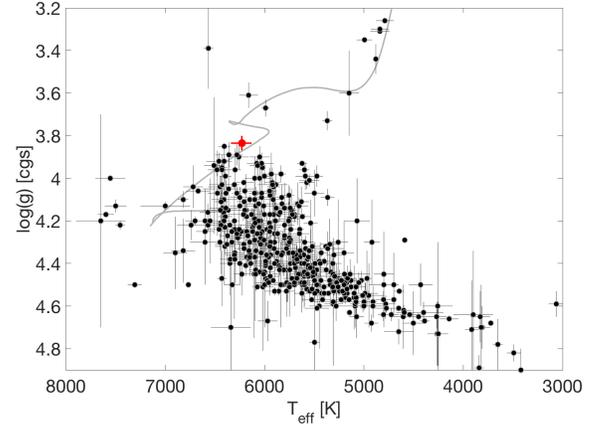}
\caption{Surface gravity and effective
temperature of the hosts of transiting
giant planets (similar to an H-R diagram).
The position of \target\ (red) falls near the edge of the occupied region of parameter space. The solid
gray line is the best-fitting MIST stellar mass track.
Data were obtained from the NASA Exoplanet Archive \citep{Akeson2013} on 2018 September 15.
\label{fig:loggteff}}
\end{figure}

The large size of \targetb\ might be connected to the evolutionary state of the host star \citep{Grunblatt2017}. \figr{loggteff} shows the location of \target\ in the space
of surface gravity and effective temperature.
\target\ is slightly evolved, with a relatively low surface gravity. As a star evolves, its luminosity increases,
which also increases the flux of radiation impinging
on any planets.
If giant planets are ``inflated'' by intense
stellar radiation, as has long been proposed,
then the larger-than-usual size of \targetb\ suggests that the evolutionary timescale of the star is slower than the inflationary timescale of the planet.

\target\ will exhaust its hydrogen fuel in $\sim0.5\,{\rm Gyrs}$, which may have ramifications for the survival of the planet. The apparent paucity of hot Jupiters orbiting evolved stars \citep{Johnson2007} has been interpreted
as a consequence of tidal destruction \citep{Villaver2009, Schlaufman2013}.
Tides raised on the star by the planet cause the planet to
transfer angular momentum to the star, a process that
is thought to accelerate rapidly as the star grows in size.
However, the timescale for this process is
unknown, with an uncertainty spanning several
orders of magnitude.

The recent discoveries of close-in gas giants around subgiants (e.g. \citealt{Van2016}), or even red giants (e.g. \citealt{Jones2018}) suggest that the lifetimes of hot Jupiters in those systems may not as short as we thought. Alternatively, as predicted by \citet{Stephan2018}, the eccentric Kozai-Lidov mechanism in a binary stellar system can drive a longer period Jupiter migrate inward during the post-main sequence phase. It seems that  with the distant stellar companion, \targetb\ is consistent with this scenario. A likely prediction from this scenario is that \targetb\ may have a non zero stellar obliquity (the angle between the orbital axis
and the stellar spin axis), which can be tested in future observations. From the stellar and planetary parameters we obtained, we predict that the Rossiter-Mclaughlin effect with have an RV semiamplitude of $10.5\,{\rm m\,s^{-1}}$ \citep[e.g.,][]{Winn2005, Gaudi2007, Albrecht2012, Wang2018}.

However, it is impossible to draw any firm conclusions
until we can measure the occurrence rates of such planets
using a homogenoeous data set.
The \textit{TESS} survey should eventually provide
the opportunity to perform such a study,
by detecting thousands of new planets orbiting a wider
variety of stars than were observed in the {\it Kepler} mission.

\acknowledgments
We thank Yanqin Wu, Smadar Naoz, Lars A. Buchhave, Bonan Pu, Alexander P. Stephan, Sabarni Basu, Beibei Liu, and Jorge Lillo-Box for their insights.
J.N.W. and S.W. thanks the Heising-Simons Foundation for their generous support.
M.N.G. acknowledges support from MIT’s Kavli Institute as a Torres postdoctoral fellow.
S.M. and A.B.D. are supported by the National Science Foundation Graduate Research Fellowship Program under Grant Number DGE-1122492.
R.B. acknowledges support from FONDECYT Post-doctoral Fellowship Project 3180246, and from the Millennium Institute of Astrophysics (MAS). 
This research is based on observations collected at 
the European Organization for Astronomical Research in the Southern Hemisphere under ESO 
programme 0101.C-0232.
This work makes use of observations from SMARTS and the LCO network.
We acknowledge the use of \textit{TESS} Alert data, which is currently in a beta test phase, from the \textit{TESS} Science Office and the \textit{TESS} Science Processing Operations Center. 
Funding for the \textit{TESS} mission is provided by NASA's Science Mission directorate.
The authors wish to recognize and acknowledge the very significant 
cultural role and reverence that the summit of Maunakea has always had 
within the indigenous Hawaiian community. We are most fortunate to have 
the opportunity to conduct observations from this mountain
This research has made use of the Exoplanet Follow-up Observation Program website, which is operated by the California Institute of Technology, under contract with the National Aeronautics and Space Administration under the Exoplanet Exploration Program.
We made use of the Python programming language \citep{Rossum1995} 
and the open-source Python packages
\textsc{numpy} \citep{vanderWalt2011}, 
\textsc{emcee} \citep{Foreman-Mackey2013}, and
\textsc{celerite} \citep{Foreman-Mackey2017}.

{\it Facilities:} 
\facility{\textit{TESS}}, 
\facility{CTIO:1.5m (CHIRON)},
\facility{ESO:3.6m (HARPS)},
\facility{Keck II (NIRC2)},
\facility{Keck I (HIRES)},
\facility{LCO:1.0m (NRES)},
\facility{FLWO:1.5m (TRES)}

\clearpage
\startlongtable
\begin{deluxetable*}{lcccc}
\tablecaption{\label{tab:params} Median values and 68\% confidence interval for \target\ planetary system.}
\tablehead{\colhead{~~~Parameter} & \colhead{Units} & \multicolumn{3}{c}{Values}}
\startdata
\smallskip\\\multicolumn{2}{l}{Stellar Parameters:}&\smallskip\\
~~~~$M_*$\dotfill &Mass (\msun)\dotfill &$1.703^{+0.075}_{-0.12}$\\
~~~~$R_*$\dotfill &Radius (\rsun)\dotfill &$2.614^{+0.080}_{-0.11}$\\
~~~~$L_*$\dotfill &Luminosity (\lsun)\dotfill &$9.25^{+0.58}_{-0.60}$\\
~~~~$\rho_*$\dotfill &Density (cgs)\dotfill &$0.134^{+0.017}_{-0.014}$\\
~~~~\logg\dotfill &Surface gravity (cgs)\dotfill &$3.835\pm0.034$\\
~~~~\teff\dotfill &Effective Temperature (K)\dotfill &$6230^{+110}_{-98}$\\
~~~~$[{\rm Fe/H}]$\dotfill &Metallicity (dex)\dotfill &$0.29^{+0.13}_{-0.24}$\\
~~~~$Age$\dotfill &Age (Gyr)\dotfill &$1.80^{+0.43}_{-0.30}$\\
~~~~$A_V$\dotfill &V-band extinction (mag)\dotfill &$0.101^{+0.050}_{-0.062}$\\
~~~~$\sigma_{SED}$\dotfill &SED photometry error scaling \dotfill &$3.6^{+3.8}_{-1.5}$\\
~~~~$\varpi$\dotfill &Parallax (mas)\dotfill &$6.79\pm0.11$\\
~~~~$d$\dotfill &Distance (pc)\dotfill &$147.2^{+2.5}_{-2.4}$\\
\smallskip\\\multicolumn{2}{l}{Planetary Parameters:}&b\smallskip\\
~~~~$P$\dotfill &Period (days)\dotfill &$3.308960\pm0.000082$\\
~~~~$R_P$\dotfill &Radius (\rj)\dotfill &$1.562^{+0.053}_{-0.069}$\\
~~~~$T_C$\dotfill &Time of conjunction (\bjdtdb)\dotfill &$2458328.68358\pm0.00035$\\
~~~~$T_0$\dotfill &Optimal conjunction Time (\bjdtdb)\dotfill &$2458338.61046\pm0.00024$\\
~~~~$a$\dotfill &Semi-major axis (AU)\dotfill &$0.05190^{+0.00075}_{-0.0012}$\\
~~~~$i$\dotfill &Inclination (Degrees)\dotfill &$84.20^{+1.1}_{-0.86}$\\
~~~~$e$\dotfill &Eccentricity \dotfill &$0.047^{+0.050}_{-0.033}$\\
~~~~$\omega_*$\dotfill &Argument of Periastron (Degrees)\dotfill &$88^{+34}_{-120}$\\
~~~~$T_{eq}$\dotfill &Equilibrium temperature$^1$ (K)\dotfill &$2132^{+37}_{-33}$\\
~~~~$M_P$\dotfill &Mass (\mj)\dotfill &$1.008^{+0.074}_{-0.079}$\\
~~~~$K$\dotfill &RV semi-amplitude (m/s)\dotfill &$96.9^{+6.1}_{-6.0}$\\
~~~~$logK$\dotfill &Log of RV semi-amplitude \dotfill &$1.986^{+0.026}_{-0.028}$\\
~~~~$R_P/R_*$\dotfill &Radius of planet in stellar radii \dotfill &$0.06144^{+0.00083}_{-0.00081}$\\
~~~~$a/R_*$\dotfill &Semi-major axis in stellar radii \dotfill &$4.27^{+0.17}_{-0.15}$\\
~~~~$\delta$\dotfill &Transit depth (fraction)\dotfill &$0.003775^{+0.00010}_{-0.000099}$\\
~~~~$Depth$\dotfill &Flux decrement at mid transit \dotfill &$0.003775^{+0.00010}_{-0.000099}$\\
~~~~$\tau$\dotfill &Ingress/egress transit duration (days)\dotfill &$0.0165\pm0.0011$\\
~~~~$T_{14}$\dotfill &Total transit duration (days)\dotfill &$0.2345^{+0.0011}_{-0.0012}$\\
~~~~$T_{FWHM}$\dotfill &FWHM transit duration (days)\dotfill &$0.21797^{+0.00062}_{-0.00058}$\\
~~~~$b$\dotfill &Transit Impact parameter \dotfill &$0.416^{+0.055}_{-0.078}$\\
~~~~$\rho_P$\dotfill &Density (cgs)\dotfill &$0.330^{+0.046}_{-0.036}$\\
~~~~$logg_P$\dotfill &Surface gravity \dotfill &$3.012^{+0.041}_{-0.039}$\\
~~~~$\fave$\dotfill &Incident Flux (\fluxcgs)\dotfill &$4.68^{+0.32}_{-0.28}$\\
~~~~$T_P$\dotfill &Time of Periastron (\bjdtdb)\dotfill &$2458328.74^{+0.49}_{-0.35}$\\
~~~~$T_S$\dotfill &Time of eclipse (\bjdtdb)\dotfill &$2458330.330^{+0.043}_{-0.051}$\\
~~~~$ecos{\omega_*}$\dotfill & \dotfill &$-0.004^{+0.020}_{-0.024}$\\
~~~~$esin{\omega_*}$\dotfill & \dotfill &$0.036^{+0.056}_{-0.037}$\\
~~~~$M_P\sin i$\dotfill &Minimum mass (\mj)\dotfill &$1.002^{+0.074}_{-0.079}$\\
~~~~$M_P/M_*$\dotfill &Mass ratio \dotfill &$0.000572^{+0.000042}_{-0.000037}$\\
~~~~$d/R_*$\dotfill &Separation at mid transit \dotfill &$4.12^{+0.30}_{-0.36}$\\
\smallskip\\\multicolumn{2}{l}{Wavelength Parameters:}&TESS\smallskip\\
~~~~$u_{1}$\dotfill &linear limb-darkening coeff \dotfill &$0.244^{+0.031}_{-0.032}$\\
~~~~$u_{2}$\dotfill &quadratic limb-darkening coeff \dotfill &$0.220^{+0.044}_{-0.042}$\\
~~~~$A_D$\dotfill &Dilution from neighboring stars \dotfill &$0.209\pm0.020$\\
\smallskip\\\multicolumn{2}{l}{Telescope Parameters:}&Chiron&HARPS&TRES\smallskip\\
~~~~$\gamma_{\rm rel}$\dotfill &Relative RV Offset (m/s)\dotfill &$-2.9^{+4.7}_{-5.2}$&$9.9^{+9.8}_{-10.}$&$97.8^{+8.5}_{-8.8}$\\
~~~~$\sigma_J$\dotfill &RV Jitter (m/s)\dotfill &$14.8^{+5.5}_{-4.1}$&$25.6^{+15}_{-8.2}$&$24.5^{+11}_{-8.4}$\\
\enddata
\label{tab:HD202772.harps+chiron+tres_nopt.}
\tablenotetext{1}{Assuming zero albedo and full heat distribution from day to night hemispheres.}
\end{deluxetable*}



\end{document}